\documentstyle[10pt,aaspp4] {article}

\begin{document}

\noindent
{\it Conference Highlights}

\begin{center}


\title{\large \bf Toward a New Millennium in Galaxy Morphology\footnote{Conference was held in Johannesburg, South Africa from 13-18 September 1999. Proceedings will be edited by D.L Block, I. Puerari, A. Stockton and D. Ferreira and 
published in a special edition of {\it Ap\&SS }}} 

\end{center}

\medskip


   The most basic process in any observational science is taxonomy - the
classification of objects by their natural relationships.   For almost a
century, the Hubble classification has been the dominate morphological
paradigm, but it appears  new ideas and systems are needed.  Many galaxies at
high redshift as well as a fraction of galaxies in the nearby universe are
unclassifiable in the Hubble sequence.  Some nearby examples 
are low surface brightness galaxies, including dwarf spheroidals, and galaxies
in clusters, especially large cD galaxies.  With so much complexity in
the galaxy population, new systems and ideas are needed.  

   We can ask at the turn of the century if optical images 
contain enough information to adequately classify galaxies.
Perhaps other wavelengths, such as the infrared, UV, radio and sub-mm are
also needed to fully cover the basic features of extragalactic systems.
Spectroscopy of a galaxy, including kinematic information, or HI line 
profiles might also be better morphological indicators than a galaxy's image.

   To address these issues, over 90 astronomers clustered in
Johannesburg, South Africa between 13-18 September to discuss the future
of galaxy morphology, and how classification relates to galaxy evolution
and formation.  Over 60 talks and a number of posters were presented during
the meeting. 

  Several new methods of classifying galaxy images were proposed at the 
conference. These new methods are in many ways different from the 
qualitative Hubble system. Not only are they quantitative in nature, with 
parameters such as asymmetry, concentration of light, Fourier components, 
measures of bar strengths, and automatic neural networks, but most do not 
reproduce the Hubble sequence, but attempt to modify or replace it (Abraham, 
Bershady, Conselice, Frei, Takamiya, Windhorst).   Physical 
morphological systems based on comparing the form of a galaxy with its
spectral-type look promising; these are similar to the ideas proposed by 
W.W. Morgan in the 1950s.  There is still however, no all encompassing system
that can relate all galaxies in different environments.

    High redshift galaxies, such as those seen in the Hubble Deep
Field (HDF) have brought about this new interest in morphology.  Although, most
of the HDF, and medium deep survey (MDS) images of distant galaxies are
sampling rest frame ultraviolet morphologies, the most disturbed galaxies
remain disturbed when viewed with NICMOS on HST.   Therefore, these galaxies
are intrinsically different from nearby ones; their morphologies are
not the result of morphological k-corrections.  We are however, only sampling
the young to middle aged stars when viewing galaxies in optical/near infrared
wavelengths.  Limiting ourselves to these wavelengths, we are missing
a large fraction of the light output from galaxies.    The effects of dust
on the morphology of galaxies, and much of the gas is being missed.

  Our view of the dusty universe is provided by ISO, IRAS, SCUBA and
other instruments that probe long-ward of 5$\mu$m (Cesarsky, Laurent,
Mirabel, Sanders, Sauvage, Scoville).   Not only does a significant
fraction of the light emitted from galaxies occur at these wavelengths, but
because of dust effects several features can only be seen by
examining galaxies at these wavelengths.  Hidden star formation in mid
ISOCAM images of nearby galaxies, prove that dust is a significant factor
affecting the morphology of galaxies.    Luminous and ultra-luminous infrared
galaxies are probably major galaxy mergers, with their optical morphology 
dominated by dust absorption.    SCUBA sources
seen at redshifts z $<$ 6, are probably analogs to nearby ULIGs and possibly
are progenitors of ellipticals with L $> 10^{12}$ L$_{\odot}$.  Many of these
questions will be answered with the next generation of sub-mm instrumentation
(Combes).

  Using near infrared images, particularly in the K band is useful for
determining the underlying structure of the old stellar populations in
galaxies.  Bars and other morphological features are more easily detected
in the NIR.  The spiral structure of galaxies, specifically the pitch
angle, changes between optical and NIR wavelengths (Block). 
Using Fourier analysis, three principle 
NIR spiral galaxy morphologies are found.  The  
pitch angle of the spiral arms in these families, viewed in the NIR 
correlates with the shape of velocity rotation profiles (Block), an effect 
not seen in optical wavelengths.

   Using dust corrected rest frame optical fluxes, infrared, sub-mm fluxes 
from high redshift galaxies and the sub-mm background, it is possible to 
obtain an estimate of the
star formation history of the universe (Lagache, Thompson).  
Determining the star formation rate in a galaxy is tricky, and estimates
of star formation rates are based on dust correction assumptions.
Although, we do not know if the dust in high redshift galaxies behaves the 
same as dust in nearby galaxies.  To paraphrase Mayo Greenberg, "Galaxies
evolve, stars evolve, so why not dust?".  Additionally, we
do not know the full effects of stellar winds, SNe, magnetic fields, and
possibly other feedback mechanisms (Chu) to say nothing of the effects of 
black holes and AGN phenomenon on the morphology of galaxies.
 
   The masses of black holes correlate with the masses of their host
galaxy's spheroidal components (Macchetteo) suggesting a possible feedback 
mechanism and related formation.  
AGNs might also play an important role in star formation in galaxies.
The space density of AGNs is related to the star formation history of
the universe (Miley).  Jet outflows are an important component for
the formation of stars in galaxies, and hence could have a significant effect 
on their morphologies.

   These physical mechanisms can change how a galaxy looks in the
general, but perhaps more detailed features deserve a closer look.  Rings, 
bars, and globular clusters 
all hold clues to galaxy formation.  Rings (Buta) can be located at 
resonances, are sites of star formation and are potentially
triggered by bars, or interactions with other galaxies.  Bars also produce
spiral arms, star formation, and drive secular evolution in spiral
galaxies.   The bar frequency among spiral galaxies in the local population 
seen in the near infrared is at about 60\% (Eskridge, Knapen).  The bar 
fraction 
decreases at higher redshifts (Abraham), suggesting that bars might form
later in the evolution of galaxies.   The specific frequency and properties
of globular clusters, especially  metallicity distributions, can also be 
used to determine the evolutionary history of galaxies (B. Elmegreen).

   An overlooked feature for morphological studies of galaxies is their gas 
content.  While some forms of ionized gas can be seen at optical 
wavelengths,  most gas is detectable in the radio.  HI 21cm lines 
images of galaxies are generally more extended than in the optical, and 
more commonly show signs of interaction and mergers than galaxies in 
optical light (Sancisi).   Interactions between galaxies is a critical 
aspect for their evolution, and any complete morphological system must 
include dynamical indices.   The effects of mergers and interactions on
galaxies, including the intensity of the induced star formation,  can 
vary depending on the initial orbital conditions (Horellou).

 HI gas dynamics can be used to determine the underlying physical
nature of galaxy disks, including dark matter halos.  Recent observations
suggest that dark matter halos are triaxial and possibly rotating in some
galaxies (Freeman).
These rotating dark matter halos, predicted by CDM could be responsible for
the structure of HI disks.
High resolution rotation curves also show that the luminous mass of galaxies
does not trace dark matter potentials in halos (Sofue).  Vertical
structures of edge-on disk galaxies are also inconsistent with slowly
rising rotation curves (van der Kruit). 

   Knowing the structure of galaxies naturally leads towards learning how 
galaxies 
formed and evolved.  Observationally we know that the galaxy population 
looked different in the past (Abraham, Conselice, Illingworth, Windhorst) 
but how did this evolution occur?  Evidence for dissipative collapse, where 
galaxies are created from the inside out, includes metallicity
gradients that correlate with bulge luminosity, and old bulges (Goudfrooij).  
On the other hand, properties of spirals from Sm to S0s suggest galaxies 
formed through a secular process (Pfenniger).  Additionally dynamical 
evolution of galaxies based on semi-analytical and N-body simulates using 
likely initial conditions of the universe suggest galaxies formed 
hierarchically. Although, dark matter halos in these models are too 
concentrated and fail to produce correct sizes for disk galaxies 
(Steinmetz).  There is however, observational evidence for hierarchical 
formation of galaxies at high redshifts (Illingworth).  Morphologies
of certain spirals can also change in special environments such as galaxy 
clusters through high speed galaxy interactions (Lake). 

  The observations and theory described above is beginning to reveal how
galaxies were formed, evolved and how fundamental properties relate to 
morphology.   A good taxonomy system for extragalactic objects should
classify galaxies according to characteristics that best describe
galaxy evolution; perhaps interactions, star formation history, gas content, 
and dark matter/rotation curve profiles, are among these.   Although such a 
morphological system does not presently exist, the accumulation of 
observational data and the further development of theory will suggest what 
parameters are the best for understanding the structure and evolution of 
galaxies.

\medskip


\noindent
{\it Christopher J. Conselice} 

\noindent
{\it University of Wisconsin, Madison}

\end{document}